\documentclass[a4paper,10pt,twocolumn]{article}

\newcounter{sectionc}\newcounter{subsectionc}\newcounter{subsubsectionc}
\renewcommand{\section}[1] {\vspace{14pt}\addtocounter{sectionc}{1}
\setcounter{subsectionc}{0}\setcounter{subsubsectionc}{0}\noindent 
	{\bf\thesectionc. #1}\par\vspace{8pt}}
\renewcommand{\subsection}[1] {\vspace{14pt}\addtocounter{subsectionc}{1}
   \setcounter{subsubsectionc}{0}\noindent 
   {\bf\thesectionc.\thesubsectionc. {\kern1pt \bfit #1}}\par\vspace{8pt}}
\renewcommand{\subsubsection}[1] {\vspace{14pt}
    \addtocounter{subsubsectionc}{1}
	\noindent{\thesectionc.\thesubsectionc.\thesubsubsectionc.
	{\kern1pt \it #1}}\par\vspace{8pt}}

\renewenvironment{thebibliography}[1]		
	{\small\baselineskip=11pt
	 \frenchspacing
	 \begin{list}{\arabic{enumi}.}
        {\usecounter{enumi}\setlength{\parsep}{0pt}    
	 \setlength{\leftmargin 12.7pt}{\rightmargin 0pt}
         \setlength{\itemsep}{0pt} \settowidth
	{\labelwidth}{#1.}\sloppy}}{\end{list}}

\newcounter{itemlistc}
\newcounter{romanlistc}
\newcounter{alphlistc}
\newcounter{arabiclistc}

\newcommand{\fcaption}[1]{
        \refstepcounter{figure}
        \setbox\@tempboxa = \hbox{\small Fig.~\thefigure. #1}
        \ifdim \wd\@tempboxa > 5in
           {\begin{center}
        \parbox{5in}{\small\baselineskip=11pt Fig.~\thefigure. #1}
            \end{center}}
        \else
             {\begin{center}
             {\small Fig.~\thefigure. #1}
              \end{center}}
        \fi}

\usepackage{graphicx}  
\usepackage{subfigure}
\usepackage{epsf}
\usepackage{amsmath}
\usepackage{amsfonts}
\usepackage{abstract}
\usepackage{balance}

\newcommand{\bra    }{\langle}
\newcommand{\ket    }{\rangle}
\newcommand{\Bra    }{\left\langle}
\newcommand{\Ket    }{\right\rangle}

\newcommand{\bm     }{\mbox{\boldmath$m$}}
\newcommand{\bq     }{\mbox{\boldmath$q$}}

\newcommand{\bL     }{\mbox{\boldmath$L$}}
\newcommand{\bG     }{\mbox{\boldmath$G$}}
\newcommand{\bD     }{\mbox{\boldmath$D$}}
\newcommand{\bsigma }{\mbox{\boldmath$\sigma$}}
\newcommand{\boldeta}{\mbox{\boldmath$\eta$}}
\newcommand{\btau   }{\mbox{\boldmath$\tau$}}
\newcommand{\bpsi   }{\mbox{\boldmath$\psi$}}

\newcommand{\bxi    }{\mbox{\boldmath$\xi$}}
\newcommand{\bC     }{\mbox{\boldmath$C$}}
\newcommand{\bR     }{\mbox{\boldmath$R$}}

\title{\bf \large ADAPTIVE THRESHOLDS FOR NEURAL NETWORKS WITH SYNAPTIC NOISE}
\author{\normalsize D. BOLL\'E and R. HEYLEN\\ 
\it \normalsize Institute for Theoretical Physics, Katholieke Universiteit Leuven\\
\it \normalsize Celestijnenlaan 200 D, B-3001, Leuven, Belgium\\
\it \normalsize E-mail: desire.bolle@fys.kuleuven.be / rob.heylen@fys.kuleuven.be
}
\date{\vspace{-1cm}}

\begin{document}

\twocolumn[
\maketitle
\renewcommand{\abstractname}{}
\renewcommand{\absleftindent}{2.5 pc}
\renewcommand{\absrightindent}{2.5 pc}
\begin{onecolabstract}
\baselineskip11pt \small 
The inclusion of a macroscopic adaptive threshold is studied for the retrieval dynamics of both layered feedforward and fully connected neural network models with synaptic noise. These two types of architectures require a different method to be solved numerically. In both cases it is shown that, if the threshold is chosen appropriately as a function of the cross-talk noise and of the activity of the stored patterns, adapting itself automatically in the course of the recall process, an autonomous functioning of the network is guaranteed.
This self-control mechanism considerably improves the quality of retrieval, in particular 
the storage capacity, the basins of attraction and the mutual information content.
\end{onecolabstract}
\vspace{0.5cm}
]

\baselineskip13pt

\section{Introduction}

In general pattern recognition problems, information is mostly encoded by a small fraction of bits and also in neurophysiological studies the activity level of real neurons is found to be low, such that any reasonable network model has to allow variable activity of the neurons. The limit of low activity, i.e., sparse coding is then especially interesting. 
Indeed, sparsely coded models have a very large storage capacity behaving as
$1/(a\ln a)$ for small $a$, where $a$ is the 
activity (see, e.g., \cite{W,P,Ga,Ok} and references therein). However,
for low activity the basins of attraction might become very small and
the information content in a single pattern is reduced \cite{Ok}.
Therefore, the necessity for a control of the activity of the neurons has
been emphasized such that the latter stays the same as the activity of the
stored patterns during the recall process.
This has led to several discussions imposing external constraints on the
dynamics of the network. However, the enforcement of such a constraint at every time
step destroys part of the autonomous functioning of the network, i.e., a functioning that has to be independent precisely from such external constraints or control mechanisms.
To solve this problem, quite recently  a self-control mechanism has been
introduced in the dynamics of networks for so-called diluted architectures \cite{DB98}. This self-control mechanism introduces  a time-dependent 
threshold in the transfer function \cite{DB98,BDA00}. It is determined as a 
function of both the cross-talk noise and the
activity of the stored patterns in the network, and adapts itself in
the course of the recall process. It furthermore allows to reach optimal retrieval performance both in the absence and in the presence of synaptic noise \cite{DB98,BDA00,BH04,DKTE02}. These diluted architectures contain no common ancestors nodes, in contrast with feedforward architectures. It has then been shown that a similar mechanism can be introduced succesfully for layered feedforward architectures but, without synaptic noise~\cite{BM00}. Also for fully connected neural networks, the idea of self-control has been partially exploited for three-state neurons \cite{BD00}. However, due to the feedback correlations present in such an architecture, the dynamics had to be solved approximately and again, without synaptic noise.

The purpose of the present work is twofold:
to generalise this self-control mechanism for layered architectures when synaptic noise is allowed, and to extend the idea of self-control in fully connected networks with exact dynamics and synaptic noise. In both cases it can be shown that it leads to a substantial improvement of the quality of retrieval, in particular the storage capacity, 
the basins of attraction and the mutual information content.

The rest of the paper is organized as follows. In Sections~2 and 3 the layered network is treated. The precise formulation of the layered model is given in Section~2 and the adaptive threshold dynamics is studied in Section~3. In Sections~4 and 5 the fully connected network is studied. The model set-up and its exact threshold dynamics is described in Section~4, the numerical treatment and results are presented in Section~5. Finally, Section~6 contains the conclusions.

\section{The layered model}
Consider a neural network composed of binary neurons arranged in layers, each
layer containing $N$ neurons. A neuron can take values ~$\sigma_{i}(t) \in
\{0,1\}$ where $t=1,\ldots,L$ is the layer index and~ $i=1, \ldots ,N$~
labels the neurons. Each neuron on layer $t$ is unidirectionally
connected to all neurons on layer $t+1$.
We want to memorize $p$ patterns  $\{\xi_i^\mu(t)\},
{{i=1,\ldots,N}, ~{\mu=1,\ldots,p}}$ on each layer $t$, taking the
values $\{0,1\}$. They are assumed to be
independent identically distributed random variables with respect
to $i$, $\mu$ and $t$, determined by the probability distribution
\begin{equation}
p(\xi_i^\mu (t))=a\delta(\xi_i^\mu (t)-1)+(1-a)\delta(\xi_i^\mu (t))
\label{adef}
\end{equation}
From this form we find that the
expectation value and the variance of the patterns are given by 
   $
     E[\xi_i^\mu (t)]=E[\xi_i^\mu (t)^2]=a~.
   $
Moreover, no statistical correlations occur, in fact for $\mu\neq\nu$
the covariance vanishes. 

The state $\sigma_{i}(t+1)$ of neuron $i$ on layer $t+1$ is determined
by the state of the neurons on the previous layer $t$ according to the  
stochastic rule 
\begin{equation}  
       \label{eq:stoc}
  P(\sigma_{i}(t+1)\mid \bsigma(t))
             = \frac{1}{1+e^{2(2\sigma_i(t+1)-1) \beta{h_i(t)}}}.
\end{equation}
with $\bsigma(t) = (\sigma_1(t), \sigma_2(t), \ldots, \sigma_N(t))$. The right hand side is the logistic function. The ``temperature" $T=1/\beta$ controls the stochasticity of the network dynamics, it measures the synaptic noise level~\cite{HKP91}.
Given the network state  $\bsigma(t)$ on layer $t$,
the so-called ``local field" ${h_i(t)}$ of neuron $i$ on the next layer $t+1$ is given by 
\begin{equation} 
       \label{eq:h}
   h_i(t)= \sum_{j=1}^{N} 
          J_{ij}(t)(\sigma_j(t) -a)-\theta(t) 
\end{equation} 
with $\theta(t)$ the threshold to be specified later. 
The couplings $J_{ij}(t)$ are the synaptic strengths of the interaction  
between neuron $j$ on layer $t$ and neuron $i$ on layer $t+1$. They
depend on the stored patterns at different layers according to the
covariance rule 
\begin{equation}   
       \label{eq:j}
   J_{ij}(t)=\frac{1}{N {a(1-a)}} \sum_{\mu=1}^{N} 
             (\xi_i^\mu (t+1)-a)(\xi_j^\mu (t)-a)~.
\end{equation}
These couplings then permit to store sets of patterns to be retrieved by
the layered network.

The dynamics of this network is defined as follows (see \cite{DKM}).
Initially the first layer (the input) is externally set in some fixed
state. In response to that, all neurons of the second layer update
synchronously at the next time step, according to the stochastic rule
(\ref{eq:stoc}), and so on.

At this point we remark that the couplings (\ref{eq:j}) are of
infinite range (each neuron interacts with infinitely many others) such
that our model allows a so-called mean-field theory approximation. This 
essentially means that we focus on the dynamics of a single neuron
while replacing all the other neurons by an average background local
field. In other words, no fluctuations of the other neurons are taken into
account.
In our case this approximation becomes exact because, crudely speaking, 
$h_{i}(t)$ is the sum of very many terms and a central limit theorem can
be applied \cite{HKP91}. 

It is standard knowledge by now that mean-field theory
dynamics can be solved exactly for these layered architectures 
(e.g., \cite{DKM,B04}). 
By exact analytic treatment we mean that, given the state of the
first layer as initial state, the state on layer $t$ that
results from the dynamics is predicted by recursion formulas.
This is essentially due to the fact that the     
representations of the patterns on different layers are chosen independently. 
Hence, the big advantage is that this 
will allow us to determine the  effects from self-control in an 
exact way. 

The relevant parameters describing the solution
of this dynamics are the 
{\it main overlap} of the  state of the network and the $\mu$-th
pattern, and the {\it neural activity} of the neurons   
\begin{eqnarray}
M^\mu(t)  &=& \frac{1}{N{a(1-a)}}\sum_{i=1}^N{(\xi_i^\mu(t)-a)} (\sigma_i(t) -a) \nonumber \\ && \label{M(t)} \\
q(t) &=& \frac{1}{N}\sum_{i=1}^N \sigma_i(t)~.
\end{eqnarray} 

In order to measure the retrieval quality of the recall process, we use
the mutual information function \cite{DB98,BDA00,NBP98,ST98}. In general, it
measures the average amount of information that can be received by
the user by observing the signal at the output of a channel \cite{B90,S48}.
For the recall process of stored patterns that we are discussing 
here, at each layer the process can be regarded as a channel with
input $\xi_i^\mu(t)$ and output $\sigma_{i}(t)$ such that this mutual
information function can be defined as \cite{DB98,B90}
\begin{equation}  
    \label{eq:inf}
   I(\sigma_i(t);\xi_i^\mu (t))=
       S(\sigma_i(t))-\langle S(\sigma_i(t)|\xi_i^\mu (t))\rangle
                                 _{\xi^{\mu}(t)}
\end{equation}
where  $S(\sigma_i(t))$ and $S(\sigma_i(t)|\xi_i^\mu (t))$ are the entropy
and the conditional entropy of the output, respectively
\begin{eqnarray} 
      \label{eq:en}
  S(\sigma_i(t))&=& -\sum_{\sigma_i} p(\sigma_i(t))\ln[p(\sigma_i(t))]\quad \\ 
      \label{eq:enc}
  S(\sigma_i(t)|\xi_i^\mu (t))&=&
       -\sum_{\sigma_i} p(\sigma_i(t)|\xi_i^\mu (t)) \nonumber \\
                &&\times \ln[p(\sigma_i(t)|\xi_i^\mu (t))]~.
\end{eqnarray}
These information entropies are peculiar to the probability 
distributions of the output. 
The quantity $p(\sigma_i(t))$ denotes the probability distribution for the
neurons at layer $t$ and $p(\sigma_i(t)|\xi_i^\mu (t))$ indicates the
conditional probability that the $i$-th neuron is in a state
$\sigma_i(t)$ at layer $t$ given that the $i$-th site of the 
pattern to be  retrieved is $\xi_i^\mu (t)$.
Hereby, we have assumed that the conditional probability of all the
neurons factorizes, i.e.,
 $p(\{\sigma_i(t)\}|\{\xi_i(t)\})=\prod_j p(\sigma_j(t)|\xi_j(t))$, which is a
consequence of the mean-field theory character of our model explained
above. We remark that a similar factorization  has also been
used in Schwenker et al.~\cite{SSP96}.

The calculation of the different terms in the expression (\ref{eq:inf})
proceeds as follows. Because of the mean-field character of our model the following formulas hold for every neuron $i$ on each layer $t$. Formally writing (forgetting about the pattern index $\mu$) $\langle O \rangle
\equiv \langle \langle O \rangle_{\sigma|\xi} \rangle_{\xi}=
\sum_{\xi} p(\xi) \sum_{\sigma} p(\sigma|\xi) O $ for an arbitrary
quantity $O$ the conditional probability can be obtained in a rather
straightforward way by using the complete knowledge about the system:
$\langle \xi \rangle=a, \, \langle \sigma \rangle=q, \,
\langle (\sigma-a)( \xi-a) \rangle=M, \, \langle 1 \rangle=1$.

The result reads
\begin{eqnarray}
    p(\sigma|\xi)&=&[\gamma_0+(\gamma_1-\gamma_0)\xi]\ \delta(\sigma-1) \nonumber \\
&& + [1-\gamma_0-(\gamma_1-\gamma_0)\xi]\ \delta(\sigma)
\end{eqnarray}
where $\gamma_0=q-aM$ and $\gamma_1=(1-a)M+q$, and where the $M$
and $q$ are precisely the relevant parameters (\ref{M(t)}) for large $N$.
Using the probability distribution of the patterns we obtain
\begin{equation}
   p(\sigma)=q\delta(\sigma-1)+(1-q)\delta(\sigma)~.
\end{equation}   
Hence the entropy (\ref{eq:en}) and the conditional entropy
(\ref{eq:enc})  become
\begin{eqnarray}
           S(\sigma)=&-&q\ln q -(1-q)\ln(1-q) \\
           S(\sigma|\xi)=&-&[\gamma_0+(\gamma_1-\gamma_0)\xi]
	             \ln[\gamma_0+(\gamma_1-\gamma_0)\xi] 
		     \nonumber \\
                &-&[1-\gamma_0-(\gamma_1-\gamma_0)\xi] \nonumber \\
                     &&\times \ln[1-\gamma_0-(\gamma_1-\gamma_0)\xi]~.
\end{eqnarray}		      
By averaging the conditional entropy over the pattern $\xi$ we finally get
for the mutual information function (\ref{eq:inf}) for the layered model
\begin{eqnarray}
     \label{eq:Ifin}
     I(\sigma;\xi) \!\!\!&=&\!\!\! -q\ln q -(1-q)\ln(1-q) \nonumber \\
          &+&a[\gamma_1\ln\gamma_1+(1-\gamma_1)\ln(1-\gamma_1)]
	  \nonumber\\
	     &+&(1-a)[\gamma_0\ln\gamma_0+(1-\gamma_0)\ln(1-\gamma_0)]~. \nonumber \\
\end{eqnarray}

\section{Adaptive thresholds in the layered network}

It is standard knowledge (e.g., \cite{DKM})
that the synchronous dynamics for layered  architectures 
can be solved exactly following the method based upon a signal-to-noise
analysis of the local field (\ref{eq:h}) (e.g., \cite{Ok,B04,A77,AM88} and references therein). 
Without loss of generality we focus on the recall of one pattern, say
$\mu=1$, meaning that only $M^1(t)$ is macroscopic, i.e., of order $1$
and the rest of the patterns causes a cross-talk noise at each 
step of the dynamics.

We suppose that the initial state of the network model $\{\sigma_i(1)\}$ 
is a collection of independent identically distributed random variables, with average and variance given by 
$
      E[\sigma_i(1)]=E[(\sigma_i(1))^2]=q_0~.
$
We furthermore assume that this state is correlated with only
one stored pattern, say pattern $\mu=1$, such that
 $  \mbox{\rm Cov}(\xi_i^\mu (1),\sigma_i(1))=\delta_{\mu,1}~M_0^1~{a(1-a)}~.$

Then the full recall proces is described by \cite{DKM,B04}

\begin{eqnarray} 
M^1(t+1) &=&  \frac{1}{2}\int{\cal D} x \left(\tanh\left[\beta F_1 \right] + \tanh\left[\beta F_2\right] \right)
\nonumber \\ && \label{eq:a}\\
q(t+1) &=& aM^1(t+1) \nonumber \\
&&+\frac{1}{2} \left( 1+\int{\cal D} x \tanh\left[\beta F_2\right] \right)
\label{eq:b}\\
D(t+1) &=& Q(t+1) \nonumber\\
&& \hspace*{-1 cm}+ \frac{\beta}{2}\left\{1-a\int{\cal D} x\tanh^2[\beta F_1] \right. \nonumber \\
&& \hspace*{-1 cm} - \left. (1-a)\int{\cal D} x\tanh^2 [\beta F_2]\right\}^2 D(t) \qquad
\label{eq:c}
\end{eqnarray}  
with
\begin{eqnarray}
F_1 &=& (1-a)M^1(t)-\theta(t)+\sqrt{\alpha D(t)}\,x \\
F_2 &=& -aM^1(t)-\theta(t)+ \sqrt{\alpha D(t)}\,x
\end{eqnarray}
and $\alpha= p/N$, ${\cal D} x$ is the Gaussian measure ${\cal D} x= dx
(2\pi)^{-1/2}\exp(-x^2/2)$, where $Q(t)=[(1-2a)q(t)+a^2]$ and where $D(t)$ contains the influence of the cross-talk noise caused by the patterns $\mu>1$. 
As mentioned before,  $\theta(t)$ is an adaptive threshold that has to be chosen. 

In the sequel we discuss two different choices and both will be compared for
networks with synaptic noise and various activities.  Of course, it is known that the quality of the recall process is influenced by the cross-talk noise. An idea is then to introduce a threshold that adapts itself autonomously in the course of the recall process and that counters, at each layer, the cross-talk noise. This is the self-control method proposed in \cite{DB98}. This has been studied for layered neural network models  without
synaptic noise, i.e., at $T=0$, where the rule (\ref{eq:stoc}) reduces  to the deterministic form 
  $  \sigma_i(t+1)=\Theta({h_i(t)}) $
with $\Theta(x)$ the Heaviside function taking the value $\{0,1\}$.
For sparsely coded models, meaning that the pattern activity $a$ is very
small and tends to zero for $N$ large, it has been found \cite{BM00} that 
\begin{equation}   
    \label{eq:thr}
     \theta(t)_{sc}= c(a)\sqrt{\alpha D(t)}, \quad  c(a)=\sqrt{-2\ln a}
\end{equation}
makes the second term on the r.h.s of Eq.(\ref{eq:b}) at $T=0$, asymptotically vanish
faster than $a$ such that $q \sim a$. 
It turns out that the inclusion of this self-control threshold  considerably improves the quality of retrieval, in particular the storage capacity, the basins of attraction and the information content.

\begin{figure}[htp]
\centering
\includegraphics[width=.45\textwidth]{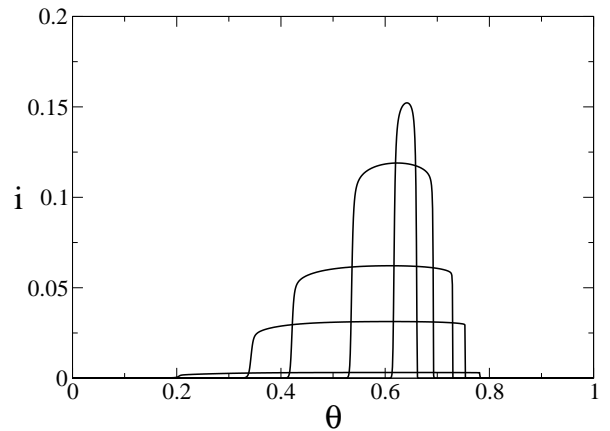}
\caption{ The information $i=\alpha I$ as a function of $\theta$ for $a=0.005$, $T=0.1$ and several values of the load parameter $\alpha=0.1,1,2,4,6$ (bottom to top)}
\label{fig:infoth}
\end{figure}

The second approach chooses a threshold by maximizing the information
content, $i=\alpha I$ of the network (recall Eq.~(\ref{eq:Ifin})). This function  depends on $M^1(t)$, $q(t)$,  $a$, $\alpha$ and $\beta$. The
evolution of $M^1(t)$ and of $q(t)$ (\ref{eq:a}), (\ref{eq:b}) depends on the specific choice of the threshold through the local field (\ref{eq:h}). We consider a
layer independent threshold $\theta(t)=\theta$  and calculate the value of 
(\ref{eq:Ifin}) for fixed $a$, $\alpha$, $M_0^1$, $q_0$ and $\beta$. The optimal 
threshold, $\theta=\theta_{opt}$, is then
the one for which the mutual information function is maximal.
The latter is non-trivial
because it is even rather difficult, especially in the limit of sparse
coding, to choose a threshold interval by hand such that $i$ is 
non-zero. The computational cost will thus be larger compared to the one of the self-control approach. To illustrate this we plot in Fig.~\ref{fig:infoth} the information content $i$ 
as a function of $\theta$  without self-control or a priori optimization, for $a=0.005$ and different values of $\alpha$.
\vspace*{0.5cm}
For every value of $\alpha$, below its critical value, there is a range for the threshold 
where the information content is different from zero and hence, retrieval is possible. This
retrieval range becomes very small when the storage capacity approaches its
critical value $\alpha_c=6.4$. 

Concerning then the self-control approach, the next problem to be posed in analogy with the case without synaptic noise is the following one. Can one determine a form for
the threshold $\theta(t)$ such that the integral in the second term on the r.h.s of Eq.(\ref{eq:b}) at $T \neq 0$ vanishes asymptotically faster than $a$? 

In contrast with the case at zero temperature where due to the simple form of the transfer function, this threshold could be determined analytically
(recall Eq.~(\ref{eq:thr})), a  detailed study of the asymptotics of the 
integral in Eq.~(\ref{eq:b}) gives no satisfactory analytic solution. 
Therefore, we have designed a systematic numerical procedure through the
following steps: 
\begin{itemize}
\item Choose a small value for the activity $a'$.
\item Determine through numerical integration the threshold $\theta'$
such that 
\begin{equation}
\int_{-\infty}^{\infty} 
     \frac{dx \,\,e^{-x^2/ 2 \sigma^2}}{\sigma \sqrt{2\pi }} 
 \Theta (x- \theta) \leq a' \quad \mbox{for} \quad \theta > \theta' 
\end{equation}
for different values of the variance $\sigma^2={\alpha D(t)}$.
\item Determine as a function of  $T=1/\beta$, the value
for $\theta'_T$ such that for $\theta > \theta' +\theta'_T$
\begin{equation}
 \int_{-\infty}^{\infty} 
     \frac{dx \,\,e^{-y^2/ \sigma^2}}{2 \sigma \sqrt{2\pi }} 
  [1+ \tanh[\beta (x- \theta)]] \leq a'
\end{equation}
\end{itemize}
The second step leads precisely to a threshold having the form of Eq.~(\ref{eq:thr}). The 
third step determining the temperature-dependent part $\theta'_T$ leads to the final 
proposal
\begin{equation}
    \theta_{t}(a,T)=\sqrt{-2 \ln (a)\alpha D(t)} - \frac12 \ln(a) T^2.
     \label{threstemp}
\end{equation}
This dynamical threshold is again a
macroscopic parameter, thus no average must be taken over the microscopic
random variables at each step $t$ of the recall process.

\begin{figure}[htp]
\centering
\includegraphics[width=.45\textwidth]{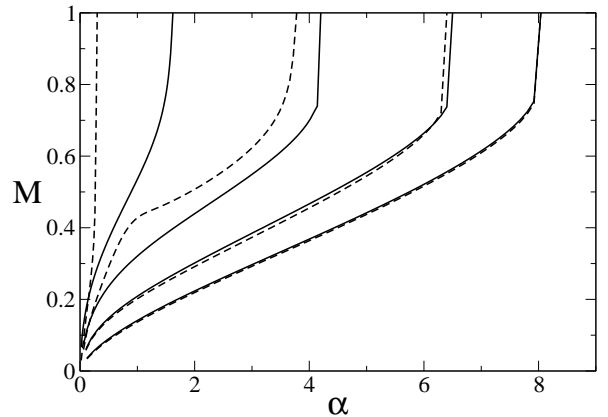}
\caption{The basin of attraction as a function of $\alpha$ for $a=0.005$
and $T=0.2, 0.15, 0.1, 0.05$ (from left to right) with 
(full lines) and without (dashed lines) the $T$-dependent part 
in the threshold (\ref{threstemp}).}
\label{fig:basinsT}
\end{figure}

\begin{figure}[ht]
\centering
\includegraphics[height=4cm,width=7.8cm]{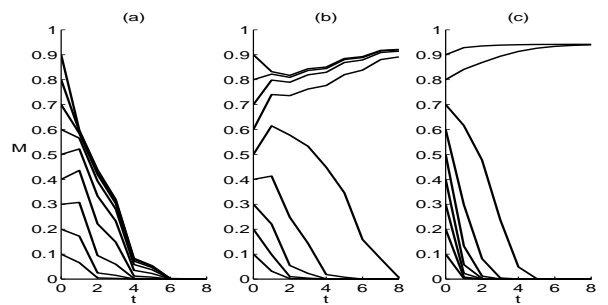}
\caption{The evolution of the main overlap $M(t)$ for
several initial values $M_0$ with $T=0.2,~q_0=a=0.005,~\alpha=1$
for the self-control model (\ref{threstemp}) without (a) and with $T$-dependent
part (b) and for the optimal threshold model (c).}
\label{fig:evolM}
\end{figure}

\begin{figure}[ht]
\centering
\includegraphics[width=.45\textwidth]{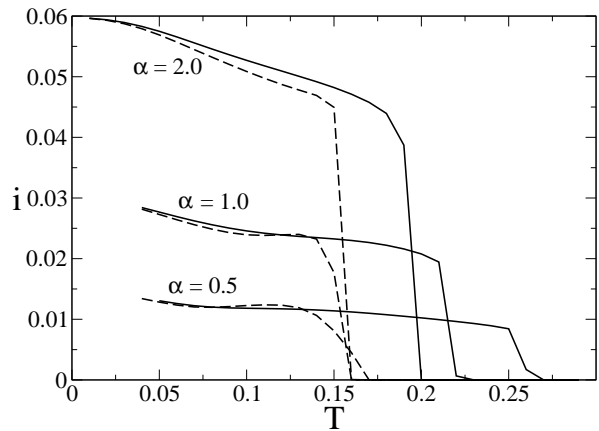}
\caption{The information content $i=\alpha I$ as a function of $T$ for several
values of the loading $\alpha$ and $a=0.005$ with (full
lines) and without (dashed lines) the $T$-correction 
in the threshold.}
\label{fig:infoT}
\end{figure}

We have solved these self-controlled dynamics,
Eqs.(\ref{eq:a})-(\ref{eq:c}) and (\ref{threstemp}), for our model with
synaptic noise, in the limit of sparse coding, numerically. In
particular, we have studied in detail the influence
of the $T$-dependent part of the threshold. Of course,
we are only interested in the retrieval solutions with $M>0$ (we forget about the index $1$) and
carrying a non-zero information~$i=\alpha I$. 
The important features of the solution are illustrated, for a typical value of $a$ in Figs.~\ref{fig:basinsT}-\ref{fig:infoT}. 
In Fig.~\ref{fig:basinsT} we show
the basin of attraction for the whole retrieval phase for the model with
threshold (\ref{eq:thr}) (dashed curves) compared to 
the model with the noise-dependent threshold (\ref{threstemp})
(full curves). We see that there is no clear improvement for low $T$
but there is a substantial one for higher $T$.
Even near the border of critical storage the results are still improved
such that also the storage capacity itself is larger. 

This is further illustrated in Fig.~\ref{fig:evolM} where we compare
the evolution of the retrieval overlap $M(t)$ starting from several initial
values, $M_0$, for the model without (Fig. \ref{fig:evolM} (a)) and with 
(Fig. \ref{fig:evolM} (b)) the $T$-correction in the threshold and for the optimal threshold model (Fig. \ref{fig:evolM} (c)). Here this 
temperature correction is absolutely crucial to guarantee retrieval, i.e., $M \approx 1$. It really makes the difference between retrieval and non-retrieval in the model. 
Furthermore, the model with the self-control threshold with noise-correction has even a wider basin of attraction than the model with optimal threshold.

In Fig.~\ref{fig:infoT} we plot the information content $i$ as a
function of the temperature for the self-control dynamics with the 
threshold (\ref{threstemp}) (full curves), respectively (\ref{eq:thr})
(dashed curves). We see that a substantial improvement of the
information content is obtained.  
\vspace*{0.5cm}

\begin{figure}[ht]
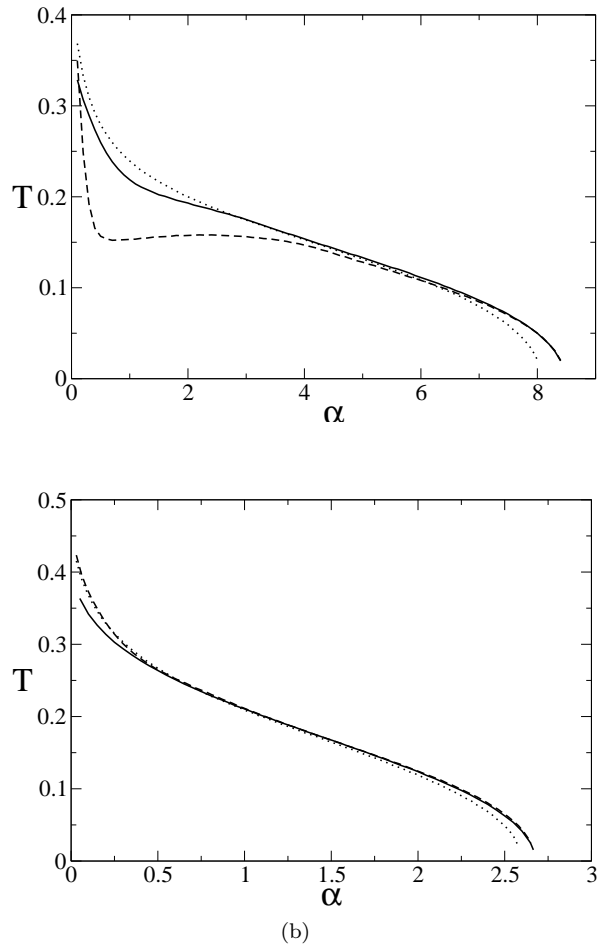

       \centering   
       \subfigure[]{
             \label{fig:phases1}
                 \includegraphics[width=.45\textwidth]{fig5a.eps}}
	 \vspace*{0.6cm}
	\subfigure[]{
	\label{fig:phases2}
             \includegraphics[width=.45\textwidth]{fig5b.eps}}
\caption{Phases in the $T-\alpha$ plane for $a=0.005$ (a) and $a=0.02$ (b) with (full
line) and without (dashed line) the temperature correction 
in the self-control threshold and with optimal threshold (dotted line).}
  \label{fig:phases}
\end{figure}

Finally we show in Fig.~\ref{fig:phases} a $T-\alpha$ plot for $a=0.005$ (a) and $a=0.02$ (b) with (full line) and without (dashed line) noise-correction 
in the self-control threshold and with optimal threshold (dotted line).
These lines indicate two phases of the layered model: below the lines our model allows recall, above the lines it does not. For $a=0.005$ we see that the $T$-dependent term in the self-control threshold leads to a big improvement in the region for large noise and small loading and in the region of critical loading. For $a=0.02$ the results for the self-control threshold with and without noise-correction and those for the optimal thresholds almost coincide, but we recall that the calculation with self-control is autonomously done by the network and less demanding computationally. 

In the next Sections we want to find out whether this self-control mechanism also works in the fully connected network for which we work out the dynamics in the presence of synaptic noise in an exact way. We start by defining the model and describing this dynamics.

\section{Dynamics of the fully connected model}

As before, the network we consider consists of $N$ binary neurons $\sigma_i \in \{0,1\}, i=1\ldots N$ but the couplings $J_{ij}$ between each pair of neurons $\sigma_i$ and $\sigma_j$ are now given by the following rule 
\begin{equation}
J_{ij} = \sum_{\mu = 1}^p (\xi_i^\mu -a)(\xi_j^\mu -a) \label{Jij}
\end{equation}
The local field is now determined by
\begin{equation}
h_i(\bsigma, t) =\frac{1}{a(1-a) N} \sum_{{j}=1}^{N} J_{ij} \sigma_j(t) + \theta (\bq) \label{localfields}
\end{equation}
The threshold is represented by the function $\theta$ and, based upon the results obtained in the previous sections and in \cite{BD00} we have chosen this to be a function of the mean activity $q$ of the neurons.

In order to study the dynamics of this model we need to define the transition probabilities for going from one state of the network to another. For each neuron at time $t+1$, $\sigma_i(t+1)$, we have the following stochastic rule (compare (\ref{eq:stoc}))
\begin{equation}
P(\sigma_i(t+1)| \bsigma(t)) = \frac{\exp (-\beta \epsilon(\sigma_i(t+1)| \bsigma(t))}
{\sum_s \exp (-\beta \epsilon(s | \bsigma(t))} \label{transprob}
\end{equation}
where 
\begin{equation}
\epsilon(\sigma_i(t+1)| \bsigma(t)) = - \sigma_i(t+1) h_i(\bsigma(t)) \label{eq:eps}
\end{equation}
with the local fields given by (\ref{localfields}) and 
where $\bsigma(0)$ at time $t=0$ is the known starting configuration. 

The dynamics is then described using the generating function analysis, which was introduced in \cite{MSR73} to the field of statistical mechanics and, by now, is part of many textbooks. The idea of this approach to study dynamics \cite{MSR73,coolendyn} is to look at the probability to find a certain microscopic path in time. The basic tool to study the statistics of these paths is the generating functional
\begin{eqnarray}
\lefteqn{Z[\bpsi] =} \nonumber \\
&&\hspace{-0.7cm}\Bra  \sum_{\bsigma(0) \ldots \bsigma(t)} \! \! \! \! P(\bsigma(0), \ldots, \bsigma(t)) e^{-i \sum_i \sum_{s=1}^t \psi_i(s) \sigma_i(s)} \! \Ket_{\! \bxi} \nonumber \\ && \label{eq:Z}
\end{eqnarray} 
with $P(\bsigma(0), \ldots, \bsigma(t))$ the probability to have a certain path in phase 
space
\begin{eqnarray}
\lefteqn{P(\bsigma(0), \ldots, \bsigma(t))}\nonumber \\
&=& P(\bsigma(0))\prod_{s=1}^t W[\bsigma(s-1), \bsigma(s)] \\
&=& P(\bsigma(0))\prod_{s=1}^t \prod_{i=1}^N P(\sigma_i(s)| \bsigma(s-1)) 
\end{eqnarray}
Here $W[\bsigma, \btau]$ is the transition probability for going from the configuration $\bsigma$ to the configuration $\btau$, and the $P(\sigma_i(s)| \bsigma(s-1))$ are given by (\ref{transprob}). In (\ref{eq:Z}) the average over the patterns $\bxi$ has to be taken since they are independent identically distributed random variables, determined by the probability distribution (\ref{adef}). 

One can find all physical observables by including a time-independent external field $\gamma_i(t)$ in (\ref{eq:eps}) in order to define a response fuction, and then calculating appropriate derivatives of (\ref{eq:Z}) with respect to $\psi_i(s)$ or $\gamma_i(t)$ letting all $\psi_i(t); i=1, \ldots, N$ tend to zero afterwards. For example we can write the main overlap $m(s)$ (as before we focus on the recall of one pattern), the correlation function $C(s,s')$ and the response function $G(s,s')$ as 
\begin{eqnarray}
m(s)&=&\frac{1}{a(1-a)N}\sum_i \xi_i \sigma_i(s) \nonumber \\
&=& i \lim_{\bpsi \rightarrow 0} \frac{1}{a(1-a)N}\sum_i \xi_i\frac{\delta Z}{\delta \psi_i(s)} \label{dpsi} \\
C(s,s') &=& \frac{1}{N}\sum_i \sigma_i(s) \sigma_i(s') \nonumber \\
&=& - \lim_{\bpsi \rightarrow 0} \frac{1}{N}\sum_i \frac{\delta^2 Z}{\delta \psi_i(s) \delta \psi_i(s')} \label{dpsi2}\\
G(s,s') &=& \frac{1}{N}\sum_i\frac{\delta}{\delta \gamma_i(s')} \sigma_i(s) \nonumber \\
&=& i \lim_{\bpsi \rightarrow 0} \frac{1}{N}\sum_i\frac{\delta^2 Z}{\delta \psi_i(s) \delta \gamma_i(s')} \label{dpsitheta}
\end{eqnarray}
The further calculation is rather technical, and we point the interested reader to the literature for more details (e.g.,\cite{coolendyn,BBV04}). One obtains an effective single neuron local field given by 
\begin{eqnarray}
h(s) &=&  
\frac{1}{a(1-a)} \left(\bm(s) - a \bq(s) \right) \left( \xi - a \right)
+ \theta (\bq ) \nonumber \\
&& + \alpha \sum_{s'=0}^{s-1} R(s,s')\sigma(s') + \sqrt{\alpha} \eta(s) 
\label{efffield}
\end{eqnarray}
with $\eta(s)$ temporally correlated noise with zero mean and correlation matrix $\bD$, and the retarded self-interaction $\bR$ which are given by
\begin{eqnarray}
\bD &=& (\mathbf{1} - \bG)^{-1} \bC (\mathbf{1} - \bG^\dagger)^{-1} \\
\bR &=& (\mathbf{1} - \bG)^{-1} 
\end{eqnarray}

The final result for the evolution equations of the physical observables is given by four self-consistent equations
\begin{eqnarray}
m(s) &=&  \langle \xi \sigma(s) \rangle_* \label{finalz} \\
q(s) &=&  \langle \sigma(s) \rangle_* \label{finalm}\\
C(s,s') &=& \langle \sigma(s) \sigma(s') \rangle_* \label{finalC}\\
G(s,s') &=& \beta \Bra \sigma(s) \Big[\sigma(s'+1)  -  \nonumber \right. \\
&& \left. \left(1 + e^{\beta h(\bsigma, \boldeta, s') }\right)^{-1} \Big] \Ket_* \label{finalG}
\end{eqnarray}
The average over the effective path measure and the recalled pattern $\bra \cdot \ket_*$ is given by
\begin{equation}
\langle g \rangle_* = \sum_\xi p(\xi) \sum_{\sigma(0),\ldots,\sigma(t)}
\int d\boldeta P(\boldeta)  P(\bsigma \mid \boldeta) g \label{effmeas}
\end{equation}
with $p(\xi)$ given by (\ref{adef}), $d\boldeta=\prod_{s'} d \eta(s')$ and with
\begin{eqnarray}
P(\boldeta) &=& \frac{1}{\sqrt{\det(2\pi \bD)}} \nonumber \\
 &\times&\! \! \exp \left( -\frac{1}{2} \sum_{s,s'=0}^{t-1} \eta(s) \bD^{-1}(s,s') \eta(s') \right) \nonumber \\
 && \label{Probx} \\
P(\bsigma \mid \boldeta) &=& \left(1 + m(0)(2 \sigma(0) - 1) - \sigma(0)\right) \nonumber \\
&&\times \left( \prod_{s=1}^t \frac{e^{\beta \sigma(s) h(s-1)}}{1 +  e^{\beta h(s-1)}} \right) \label{Probpath}
\end{eqnarray}
Remark that the term involving the one-time observables in (\ref{efffield}) has  the form $\left(\bm - a \bq\right)$. Therefore, in the sequel we define the main overlap  $M$ as
\begin{equation}
M = \frac{1}{a(1-a)} (\bm - a\bq) \quad \in [-1,1]
\end{equation}
The set of equations (\ref{finalz}), (\ref{finalm}), (\ref{finalC}) and (\ref{finalG}) represent an exact dynamical scheme for the evolution of the network.

To solve these equations numerically we use the  Eisfeller and Opper method (\cite{eisopper}). The algorithm these authors propose is an advanced Monte-Carlo algorithm.  Recalling equation (\ref{effmeas}) this requires samples from the correlated noise (for the integrals over $\eta$), the neurons (for the sums) and the pattern variable $\xi$. Instead of generating the complete vectors at each timestep, we represent these samples by a large population of individual paths, where each path consists of $t$  neuron values, $t$ noise values and one pattern variable. All the averages (integrations, sums and traces over probability distributions) can then be represented by summations over this population of single neuron evolutions. Because of causality, we also know that it is possible to calculate a neuron at time $s$ when we know all the variables (neurons, noise, physical observables) at previous timesteps. Also, the initial configuration at time zero is known. This gives rise to an iterative scheme allowing us to numerically solve the equations at hand.

The main idea then is to represent the average (\ref{effmeas}) over the statistics of the single particle problem, as an average over the population of single neuron evolutions. Since we did not find an explicit algorithm in the literature we think that it is very useful to write one down explicitly.
\begin{itemize}
\item Choose a large number $K$, the number of independent neuron evolutions in the population, a final time $t_f$, an activity $a$, a pattern loading $\alpha$, and an initial condition (an initial overlap, correlation, activity, ...).
\item Generate space for $K$ neuron evolutions $p_i$. Each evolution contains a pattern variable $\xi_i \in \{0,1\}$, $t_f$ neuron variables $\sigma_i(s) \in \{0,1\}$, and $t_f$ noise variables $\eta_i(s) \in \mathbb{R}, s=0 \ldots t_f, i=1 \ldots K$.
\item At time $0$, initialize the $\xi_i$ according to the distribution (\ref{adef}). Then initialize the neuron variables at time zero employing the initial condition, e.g.:
\begin{eqnarray*}
&&\text{When an initial activity is defined: }\\ 
&&\qquad P(\sigma_i(0) = 1) =  q(0) \\
&&\text{When an initial overlap is defined: }\\ 
&&\qquad P(\sigma_i(0) = \xi_i) =  M(0) 
\end{eqnarray*}
\item The algorithm is recursive. So, at time $t$ we assume that we know the neuron variables for all times $s\leq t$, the noise variables for all times $s < t$, and the matrix elements $D(s, s')$ for $s, s'<t$. We want to first calculate the noise variables at time $t$, and then the  neuron variables at time $t+1$. At timestep $t$ this can be done as follows
\begin{enumerate}
\item Calculate the physical observables $m(t)$, $q(t)$ and $C(t,s) = C(s,t)$, $s\leq t$, by summing over the population:
\begin{eqnarray}
m(t) &=&  \frac{1}{K} \sum_{i=1}^K \xi_i \sigma_i(t)  \\
q(t) &=&  \frac{1}{K} \sum_{i=1}^K \sigma_i(t)  \\
C(t,s) &=&  \frac{1}{K} \sum_{i=1}^K \sigma_i(t)\sigma_i(s)  
\end{eqnarray}
\item For $s<t$ calculate the matrix $\bL$
\begin{equation}
L(t,s) = \frac{1}{K} \sum_{i=1}^K \sigma_i(t) \eta_i(s)
\end{equation}
\item Calculate $\bG = \alpha^{-1/2}\bL \bD^{-1}$, where $\bD$ is the known noise correlation matrix from the previous timestep. Turn $\bG$ into a square matrix by adding a column of zeros to the end.
\item Calculate $\bR = (\mathbf{1} - \bG)^{-1}$ and the new $\bD = \bR \bC \bR^\dagger$
\item For each site $i$, calculate a new noise variable:
\begin{eqnarray}
\eta_i(t) &=& \frac{\zeta_i(t)}{\sqrt{\bD^{-1}(t,t)}} \nonumber \\
&&- \frac{1}{\bD^{-1}(t,t)} \sum_{s<t}\bD^{-1}(t,s)\eta_i(s) \nonumber \\
\end{eqnarray}
where all $\zeta_i(t)$ are independently chosen from a standard gaussian distribution.
\item Calculate the effective local field at each site:
\begin{eqnarray}
h_i(t) &=&  M(t) \left( \xi_i - a \right) +  \theta ( q(t) ) \nonumber \\
 && + \alpha \sum_{s\leq t} R(t,s) \sigma_i(s) + \sqrt{\alpha} \eta_i(t) \nonumber \\
 \label{algofield}
\end{eqnarray}
\item Use this local field to determine the new spin value at each site at time $t+1$:
\begin{equation}
P(\sigma_i(t+1)) = \frac{e^{\beta \sigma_i(t+1) h_i(t)}}{1 +  e^{\beta h_i(t)}}
\end{equation}
\item If $t<t_f$ increase $t$ and go to step 1. Else stop.
\end{enumerate}
\end{itemize}

This algorithm can be easily performed in a parallel way. All individual  neuron evolutions are independent of each other, and the only steps that cannot be executed in a distributed fashion are steps 3 and 4. It turns out that these two steps mostly take less than 1\% of the total calculation time.

\section{Thresholds in the fully connected network}

We have used the algorithm above to check the evolution of the overlap. The threshold function $\theta(q(t))$ appears in the local field (\ref{algofield}), and its effect on the evolution of the different physical observables can be investigated.

We take the size of the population of independent neuron evolutions $K = 10^6$. Larger population sizes can be obtained by making the algorithm parallel, but no significant differences are found.

\begin{figure}
\begin{center}
\includegraphics[width=.45\textwidth]{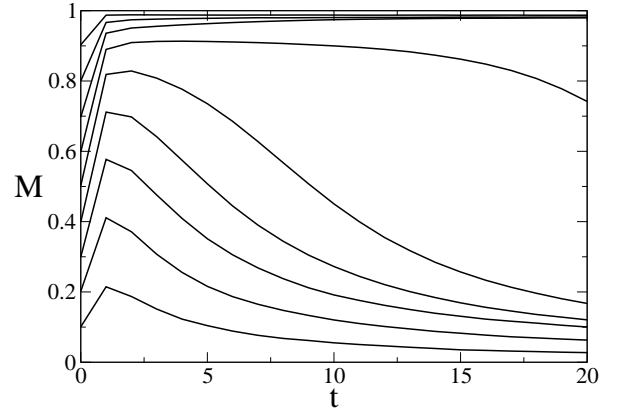}
\caption{The evolution of the overlap of the fully connected network for several initial overlaps. The system parameters are $\alpha=0.06$, $a=0.5$, $T=0.04$ and $\theta(q) = 0$.}
\label{evolution1}
\end{center}
\end{figure}

We first look at the unbiased case $(a=1/2)$ without threshold. In fig. \ref{evolution1} we plot the evolution of the overlap $M$ for several initial conditions. When the initial overlap $M_0$ is too smal there is no retrieval. This critical initial overlap separating a retrieval phase from a non-retrieval phase forms the border of the basin of attraction. 
For biased low activity networks, it is already known (e.g, \cite{Ok}) that a constant threshold $(a-1/2)$ has to be introduced in the local field eq. (\ref{localfields}) in order to guarantee a correct functioning of the network. This can easily be seen by noting that for a network where only one single pattern is stored $(h_i \rightarrow \xi_i-a)$ such that the field becomes $(1-a)$ or $(-a)$. And this lies completely asymmetric with respect to the symmetric (around the point $1/2$) transfer function eq.(\ref{transprob}). For $a \rightarrow 0$ one even finds that the probability that a neuron changes its state from zero to one becomes $1/2$.

\begin{figure}
\begin{center}
\includegraphics[width=.45\textwidth]{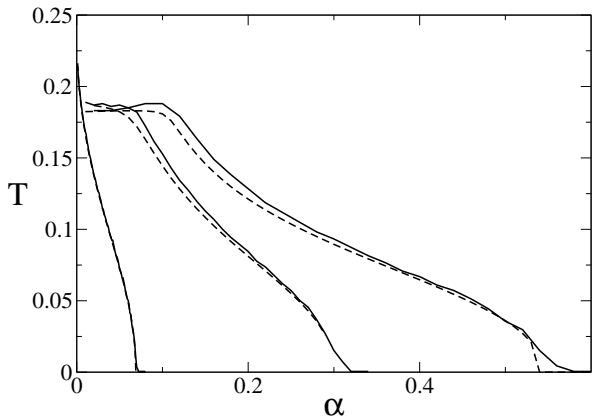}
\caption{Phases in the $T- \alpha$ plane for, from left to right, $a = 0.5$, $a=0.1$, $a=0.05$ with $\theta(q) = a - 0.5$. Solid (dashed) lines indicate the results for the dynamics (statics).}
\label{phase_diffa}
\end{center}
\end{figure}

A $T-\alpha$ plot for several values of the activity with $\theta(q) = a - 0.5$ is presented in fig. \ref{phase_diffa}.
The solid lines represent the results from the dynamics obtained by initializing the algorithm discussed in section 4 with an initial overlap $M_0=1$, and determining the temperature where this overlap has decreased below $0.4$ after $200$ timesteps. For comparison the dashed lines show the results from an equilibrium statistical mechanics calculation (e.g., \cite{horner,AGS}). As to be expected, both calculations agree. These lines indicate two phases of the fully connected model: below the lines our model allows recall, above the lines it does not.

The main question we want to address in this Section is whether we can again improve the retrieval capacities of this network architecture by introducing the self-control threshold (\ref{threstemp}). We recall that the quantity $D(t)$ occurring in this expression contains the influence of the cross-talk noise. From the signal-to-noise ratio analysis in \cite{BD00} and from statistical neurodynamics arguments (\cite{AM88}) we know that the leading term of $D(t)$ is $q(t)$. Moreover, from a biological point of view, it does not seem plausible that a network monitors the statistical quantity of the cross-talk noise. Therefore, we take $D(t)=q(t)$ in the self-control threshold in fully connected networks. 

\begin{figure}
\begin{center}
\includegraphics[width=.45\textwidth]{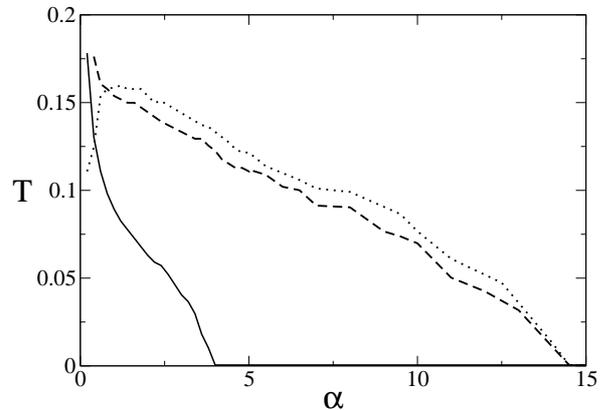}
\caption{Phases in the $T-\alpha$ plane for $a=0.005$ and several thresholds. Solid:  $\theta= a-0.5$; dashed: self-control threshold without T-correction; dotted: self-control threshold with T-correction. }
\label{phase_temp}
\end{center}
\end{figure}

\begin{figure}[ht]
\begin{center}
\includegraphics[width=.45\textwidth]{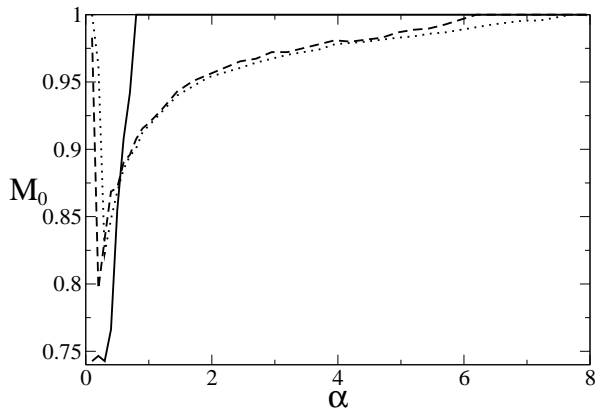}
\caption{The basin of attraction as a function of $\alpha$ for $a=0.005$ and $T=0.1$. Solid: $\theta= a-0.5$; dashed: self-control threshold without T-correction; dotted: self-control threshold with T-correction.}
\label{basin1_temp}
\end{center}
\end{figure}

\begin{figure}[ht]
\begin{center}
\includegraphics[width=.45\textwidth]{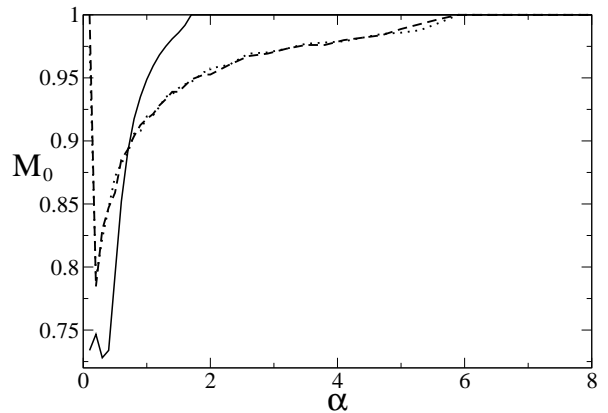}
\caption{The basin of attraction as a function of $\alpha$ for $a=0.01$ and $T=0.05$. Solid: constant $\theta= a-0.5$; dashed: self-control threshold without T-correction; dotted: self-control threshold with T-correction. }
\label{basin2_temp}
\end{center}
\end{figure}

We have then solved the generating functional analysis (\ref{finalz})-(\ref{finalG}) with the threshold
\begin{equation}
    \theta(q(t))=\sqrt{-2 \ln (a)\alpha q(t)} - \frac12 \ln(a) T^2.
     \label{threstempcon}
\end{equation}
Some typical results are shown in figs. \ref{phase_temp}-\ref{basin2_temp}. For system parameters comparable with those for the layered architecture, fig. \ref{phase_temp} clearly shows that the self-control threshold without T-correction significantly increases the retrieval region, and the temperature correction further improves the results for $\alpha$ not too small.

Looking at a fixed $T=0.1$ for this case (Fig. \ref{basin1_temp}), we furthermore notice that the self-control threshold without T-correction again significantly increases the basin of attraction. The additional temperature correction further increases this basin, and even increases the maximal achievable pattern loading $\alpha$.

\vspace*{0.5cm}
For lower temperatures (Fig. \ref{basin2_temp}) the self-control threshold still increases the basin of attraction for larger values of the pattern loading $\alpha$, but for smaller loadings the effect is diminishing. The temperature correction gives no clear improvement in this case. A similar behavior was observed for the layered architecture in fig. \ref{fig:basinsT}.
\vspace*{0.5cm}
We remark that the subtraction of ($a-1/2$) is not necessary when using the self-control method. The latter takes this into account automatically and the networks operates fully autonomously.

\section{Conclusions}
In this work we have studied the inclusion of an adaptive threshold in 
sparsely coded layered and fully connected neural networks with synaptic noise. 
We have presented an analytic form for a self-control threshold, allowing an autonomous functioning of these networks, and compared it, for the layered architecture, with an optimal threshold obtained by maximizing the mutual information which has to be calculated externally each time one of the network parameters (activity, loading, temperature) is changed.
The consequences of this self-control mechanism on the quality of the recall process have been studied.

We find that the basins of attraction of the retrieval solutions as well
as the storage capacity are enlarged. For some activities the self-control threshold even sets the border between retrieval and non-retrieval.
This confirms the considerable improvement of the quality of recall by 
self-control, also for layered and fully connected network models with synaptic noise. 

This allows us to conjecture that 
self-control might be relevant even for dynamical systems in general, when
trying to improve, e.g.,  basins of attraction.\\

\noindent
{\bf Acknowledgment}\\

\noindent 
This work has been supported by the Fund for Scientific Research- Flanders
(Belgium). \\

\noindent
{\bf References}\\

\end{document}